\documentclass[12pt]{article}
\usepackage{amssymb,graphicx}
\usepackage{amsmath,amsfonts}
\usepackage{epstopdf}
\usepackage{esint}
\usepackage{tabularx}
\usepackage{hyperref}
\usepackage{braket}
\usepackage{array}
\usepackage{subcaption}
\usepackage{bbold}
\usepackage{epsfig}
\usepackage{xcolor}
\usepackage[normalem]{ulem}

\def\e{{\rm e}}

\newcommand{\be}{\begin{equation}}
\newcommand{\ee}{\end{equation}}
\newcommand{\bea}{\begin{eqnarray}}
\newcommand{\eea}{\end{eqnarray}}
\newcommand{\bg}{\begin{gather}}
\newcommand{\eg}{\end{gather}}
\newcommand{\bseq}{\begin{subequations}}
\newcommand{\eseq}{\end{subequations}}

\begin{document}
\begin{flushright}
INR-TH-2022-011
\end{flushright}
\vspace{10pt}
\begin{center}
  {\LARGE \bf Unitarity relation and unitarity bounds\\
    for scalars with different sound speeds}\\
\vspace{20pt}
%\medskip
Y.~Ageeva$^{a,b,c,}$\footnote[1]{{\bf email:}
    ageeva@inr.ac.ru}, P. Petrov$^{a,}$\footnote[2]{{\bf email:}
    petrov@inr.ac.ru}\\
\vspace{15pt}
  $^a$\textit{
Institute for Nuclear Research of
         the Russian Academy of Sciences,\\  60th October Anniversary
  Prospect, 7a, 117312 Moscow, Russia}\\
\vspace{5pt}
$^b$\textit{Department of Particle Physics and Cosmology,\\
  Physics Faculty, M.V.~Lomonosov
  Moscow State University, \\Leninskie Gory 1-2,  119991 Moscow, Russia
  }\\
\vspace{5pt}
$^c$\textit{
  Institute for Theoretical and Mathematical Physics,\\
  M.V.~Lomonosov Moscow State University,\\ Leninskie Gory 1,
119991 Moscow,
Russia
}
    \end{center}
\vspace{5pt}

\begin{abstract}
%
%  {\bf ABSTRACT NOT EDITED}
  Motivated by scalar-tensor gravities, we
  consider a theory which contains massless scalar fields with
  different sound speeds. We derive unitarity relations
  for partial wave amplitudes of
  $2 \to 2$ scattering, with explicit formulas for contributions of
  two-particle intermediate states. Making use of these relations, we obtain
  unitarity bounds both in the most general case and in the case
  considered in literature for unit sound speed. These bounds can be used for
  estimating the strong coupling scale of a pertinent EFT. We illustrate
  our unitarity relations  by explicit calculation
  to the first non-trivial order in couplings
  in a simple model of two scalar fields
  with different sound speeds.

\end{abstract}    

\section{Introduction}

Scalar-tensor theories of gravity with non-trivial scalar kinetic terms and/or non-minimal
couplings to metric are commonly used to construct models of
inflation~\cite{Armendariz-Picon:1999hyi,Garriga:1999vw,Alishahiha:2004eh,Kobayashi:2010cm,Kobayashi:2011nu} as well as 
non-singular cosmological models such as genesis \cite{Creminelli:2010ba,Hinterbichler:2012fr,Pirtskhalava:2014esa,Nishi:2015pta,Kobayashi:2015gga,Kolevatov:2017voe}
and bounce  \cite{Qiu:2011cy,Easson:2011zy,Cai:2012va,Osipov:2013ssa,Qiu:2013eoa,Koehn:2013upa,Qiu:2015nha,Ijjas:2016tpn,Mironov:2018oec}. In these theories,
perturbations about non-trivial backgrounds often propagate with
``sound speeds'' different from
the speed of light and, moreover, perturbations of  different types 
(e.g., scalar vs tensor in the cosmological context) have different sound speeds.
Another feature is that
some constructions involve time-dependent couplings which are
dangerously large
during certain time intervals. An example is Horndeski theory~\cite{Horndeski:1974wa}
whose subclass admits
genesis and bounce with ``strong gravity in the past'' (effective Planck mass tends to zero as
$t\to -\infty$)~\cite{Ageeva:2021yik};
in this way one evades the no-go
theorem of Refs.~\cite{Libanov:2016kfc,Kobayashi:2016xpl}.

An important parameter in an effective QFT is the energy scale of strong coupling,
or, in other words, the maximum energy below which the effective QFT description
is
trustworthy. Scalar-tensor gravities, especially featuring large couplings,
are not exceptional in this regard. While the
strong coupling energy scale can often be qualitatively estimated by naive
dimensional analysis, more accurate estimates are obtained using
unitarity bounds that follow from general unitarity relations. 
This motivates us to  derive unitarity relations and
unitarity bounds in theories with different sound speeds of different perturbations.

In this paper we consider theories with several {\it scalar} fields;
theories of particles with spin can be treated in a similar
way\footnote{Bosonic perturbations with spin can often be reduced to effective
  scalar
  perturbations at the expense of the violation of Lorentz invariance, which
  occurs in non-trivial backgrounds anyway.}. 
   Also, we study theories
in {\it flat}
space-time and {\it trivial} background; this treatment is expected to be relevant also for
 non-trivial backgrounds, since  
the classical description of a background is legitimate
provided that its classical energy scale is well below the quantum strong coupling
scale, in which case the space-time dependence of the background is expected to be
negligible when  evaluating the
quantum scale.

An adequate approach to unitarity relations and unitarity bounds makes use of
Partial Wave Amplitudes (PWAs) (see, e.g.,
Refs.~\cite{Oller:2019opk,oller.190503.1,Lacour:2009ej,Gulmez:2016scm}).
We 
follow this approach in our paper. We aim at self-contained presentation and
give detailed derivation even though many steps follow closely the analyses
existing in literature. In this sense this paper may serve as a pedagogical mini-review of the
subject, with the novelty due to the fact that we consider different sound speeds of
different excitations.

This paper is organized as follows. We derive in
Sec.~\ref{sec:unit_rel} the general unitarity relations for PWAs of
$2 \to 2$ scattering, paying special attention to two-particle intermediate states.
In Sec.~\ref{sec:unit_rel} we also derive the  unitarity bounds. 
To this end,  in
Sec.~\ref{subsec:gen} we describe the class of theories we deal with.
We then study separately the cases of a pair of
distinguishable
particles in the intermediate state
(Sec.~\ref{sec:distinguishable}) and a pair of
identical particles  (Sec.~\ref{sec:identical}).
Unitarity bounds are derived in Sec.~\ref{subsec:bound}.
We give an illustrative example in Sec.~\ref{sec:unit_rel_holds}
where we explicitly check the validity of the unitarity relation to the leading non-trivial
order in a simple model of two
real scalar 
fields.
Appendix A is dedicated 
to the  time-reversal invariance and its consequence for PWAs.

\section{Unitarity relation}
\label{sec:unit_rel}

\subsection{Generalities}
\label{subsec:gen}

In this Section we proceed in the spirit of
Ref.~\cite{Oller:2019opk} and obtain the
unitarity relation for
$2\to 2$ scattering processes
in theories with scalar fields $\phi_i$
whose  sound 
speeds $u_i$ are different. Having in mind the issue of
strong coupling
energy scale, we neglect masses of these particles (if they exist).
The quadratic action reads
\begin{equation*}
  S= \sum_i S_{\phi_i} \; , \;\;\;\;
   S_{\phi_i} = \int d^4 x\left(\frac{1}{2}\dot{\phi_i}^2 - \frac{1}{2}u_{i}^2 (\vec{\nabla}\phi_i)^2\right).
\end{equation*}
The linearized equation of motion for $\phi_i$ is
\begin{equation*}
    \ddot{\phi_i} - u_{i}^2\Delta\phi_i = 0,
\end{equation*}
and its solution can be written as follows
\begin{equation*}
    \phi_i(\vec{x},t) = \int \frac{d\vec{p}_i}{(2\pi)^3}\frac{1}{\sqrt{2E_{p_i}}}\Big(a_{\vec{p}_i} 
    \e^{-iE_{p_i}t + i\vec{p}_i\vec{x}}+a_{\vec{p}_i}^{\dagger} \e^{iE_{p_i}t - i\vec{p}_i\vec{x}}\Big),
\end{equation*}
where 
\begin{equation}
\label{disp_rel}
    E_{p_i} = u_{i}p_i \; ,
\end{equation}
and the operators
$a_{\vec{p}_i}$ and $a_{\vec{p}_i}^{\dagger}$ obey the standard 
commutational relation
\begin{equation}
\label{comm_rel}
      [a_{\vec{p}_i^{\, \prime}}, a_{\vec{p}_j}^{\dagger}]
        = (2\pi)^3\delta^{(3)}(\vec{p}_i^{\, \prime}
        - \vec{p}_j )\delta_{ij}.
\end{equation}
We define one-particle state as follows:
\begin{equation*}
    \ket{\vec{p}_i\;} \equiv \sqrt{2 E_{p_i}} a_{\vec{p}_i}^{\dagger}\ket{0}  ,
\end{equation*}
so that one has the standard relation
\be
\braket{0|\phi_i (\vec{x}, t)|\vec{p\,}_j} = \e^{-iE_{p_j}t + i\vec{p}_j\vec{x}}
\delta_{ij}\; , \nonumber
\ee
while the normalization of this state is given by
\begin{equation}
\label{norm_1_part}
    \braket{\vec{p\;}^\prime_j|\vec{p\;}_i} =  (2\pi)^3 \sqrt{2E_{p_j^\prime} 2E_{p_i} } \delta^{(3)}(\vec{p}_i-\vec{p}_j^{\, \prime})\delta_{ij}.
\end{equation}
In the $i$-th  one-particle sector one has
\begin{equation*}
     \mathbb{1} = \int \frac{d^3 p_i}{(2\pi)^3 2E_{p_i}}  \ket{\vec{p\;}_i}\bra{\vec{p\;}_i}.
\end{equation*}

The $S$-matrix and $T$-matrix are related in the standard way:
\begin{equation*}
    S =  \mathbb{1} + i T,
\end{equation*}
and one extracts from $T$ the overall  $\delta$-function of
%Since the S-matrix should vanish unless the initial and final states have the same total 
4-momentum conservation:
\begin{equation}
    \label{matrix_element}
    T = (2\pi)^4 \delta^4\big(\mathcal{P}^{\mu \, \prime} - \mathcal{P}^{\mu}
    \big) M \; ,
\end{equation}
where  $\mathcal{P}^{\mu}  =\sum p^{\mu}_{in}$ and
$\mathcal{P}^{\mu \, \prime}  =\sum p^{\mu}_{out}$ are total 4-momenta of the
initial and final state, respectively.

Now, we consider an initial state 
\be
\ket{\psi,\beta} =  \sqrt{2E_{p_1}}\sqrt{2E_{p_2}} a^{\dagger}_{p_1}a^{\dagger}_{p_2}\ket{0},
\label{jun3-22-1}
\ee
with 
two particles of momenta $\vec{p}_1$ and $\vec{p}_2$, and a final state
$\ket{\psi',\beta'}$ with 
two particles of momenta
$\vec{p}_1\,'$ and $\vec{p}_2\,'$. Notation $\beta$
refers to the types of the two particles, $\beta = \{\phi_i , \phi_j \}$,
while notation $\psi$ is a shorthand for the pair of momenta,
$\psi = \{ \vec{p}_1 , \vec{p}_2\}$. Thus,
\be
\ket{\psi,\beta} =  \ket{\phi_i , \vec{p}_1}\otimes\ket{\phi_j ,\vec{p}_2} \; .\nonumber
\ee
 In eq.~\eqref{jun3-22-1} we do not explicitly indicate the type of particle to simplify formulas
and write $a_{p_1}^\dagger \equiv a^\dagger_{i \, p_1}$, etc.

 Our purpose is to derive the unitarity relation for the partial
wave amplitudes.

\subsection{Distinguishable particles}
\label{sec:distinguishable}

Let us begin with the case of distinguishable particles in
  a pair  $\beta = \{\phi_i , \phi_j \}$. 
In the next subsection we consider the case of identical particles.

The scalar product of states  $\ket{\psi',\beta'}$ and $\ket{\psi,\beta}$  is
\begin{equation}
\label{norm_2_part}
    \braket{\psi',\beta'\;|\psi,\beta\;} = (2\pi)^6 2E_{p_1}2E_{p_2}\delta^{(3)}(\vec{p}_1\;'-\vec{p}_1)\delta^{(3)}(\vec{p}_2\;'-\vec{p}_2) \delta_{\beta' \beta}\; .
\end{equation}
%(note, that Oller denote this expression as $\braket{\vec{p}\;'|\vec{p}}$, formula (2.27), where initial momenta are $\vec{p}$ and $-\vec{p}$, and final ones are $\vec{p}\;'$ and $-\vec{p}\; '$)
This follows from the
one-particle state normalization \eqref{norm_1_part}. 
In what follows we consider
the center-of-mass frame of the two-particle system.  In this frame we
denote
$\vec{p}\equiv\vec{p}_1 = - \vec{p}_2$,
$p \equiv|\vec{p}\;| = |\vec{p}_1| = |\vec{p}_2|$.
%, and
%$\mathcal{P^{\mu}}\equiv p_{1}^{\mu} + p_{2}^{\mu} = (E,0)$.
Let $\hat{\vec{p}} = \vec{p}/p$ be the unit vector along $\vec{p}$ and
$\theta, \phi$ be the corresponding angles.
We  now replace the variables $\vec{p}_1$,
$\vec{p}_2$ in \eqref{norm_2_part} 
by $\mathcal{P}^{\mu} \equiv p_1^\mu + p_2^\mu$, $\theta$ and $\phi$, where we
have in mind that in the vicinity of the center-of-mass frame one has
$\mathcal{P}^{\mu} \approx (E, 0)$, where $E= (u_{1\beta}+ u_{2\beta})p$
and $u_{1\beta}\equiv u_i$,  $u_{2\beta} \equiv u_j$
are sound speeds of the two particles in the pair
 $\beta = \{\phi_i , \phi_j \}$.
%, where $0\leq \theta\leq\pi$
%and $0\leq\phi<2\pi$.
%We denote this solid angle as
%$\hat{\vec{p}}$.
%Particle $p_1$ has sound speed $u_{1\beta}$, while $p_2$ has sound 
%speed $u_{2\beta}$. 
For the volume element we have
\begin{equation*}
    d^3\vec{p}_1d^3\vec{p}_2 = d^3\vec{\mathcal{P}} p^2 dp d\hat{\vec{p}} = \frac{p^2}{(u_{1\beta}+u_{2\beta})}d^4\mathcal{P}^{\mu} d\hat{\vec{p}} ,
\end{equation*}
which gives
%where we use dispersion relation \eqref{disp_rel}.
%Changing the variables, one can write
\begin{equation*}
\label{vol_delta}
\delta^{(3)}(\vec{p}_1\;'-\vec{p}_1)\delta^{(3)}(\vec{p}_2\;'-\vec{p}_2) \delta_{\beta\beta'} =  \frac{(u_{1\beta}+u_{2\beta})}{p^2}
\delta^{(4)}(\mathcal{P}^{\mu\, \prime} - \mathcal{P}^{\mu})
\delta^{(2)}(\hat{\vec{p}}\;'-\hat{\vec{p}})\delta_{\beta\beta'},
\end{equation*}
%That is why using \eqref{disp_rel}, \eqref{norm_2_part}
and hence
%\eqref{vol_delta}, we obtain
\begin{align}
\label{psi_psi'}
\braket{\psi',\beta'\;|\psi,\beta\;}
%&= (2\pi)^6 (2u_{1\beta}|\vec{p}_1|2u_{2\beta}|\vec{p}_2|)\delta^{(3)}(\vec{p}_1\;'-\vec{p}_1)\delta^{(3)}(\vec{p}_2\;'-\vec{p}_2) \delta_{\beta\beta'} \nonumber\\
=(2\pi)^6 \cdot 4u_{1\beta}u_{2\beta}(u_{1\beta}+u_{2\beta})\cdot
\delta^{(4)}(\mathcal{P}^{\mu \, \prime} - \mathcal{P}^{\mu})
\delta^{(2)}(\hat{\vec{p}}\;'-\hat{\vec{p}})\delta_{\beta\beta'}.
\end{align}

As the next step we introduce two-particle state of definite angular momentum
in the center-of-mass frame. The reason is that the
unitarity relations have particularly
simple form for the partial-wave
amplitudes~\cite{Oller:2019opk,oller.190503.1,Lacour:2009ej,Gulmez:2016scm} 
(PWAs). The relevant state is given by
\begin{equation}
\label{LSJ_state}
\ket{l, m,\mathcal{P}^{\mu},\beta}
= \frac{1}{\sqrt{4\pi}} \int d\hat{\vec{p}}\;Y^m_l(\hat{\vec{p}})
\ket{\psi\;,\beta},
\end{equation}
where the integration runs over unit sphere 
and $Y^m_l$ is the spherical function,
\begin{equation*}
\label{sph_func}
Y^{m}_{l}(\hat{\vec{p}}\;) =  (-1)^{\frac{|m|-m}{2}}\mbox{e}^{im\phi}
\sqrt{\frac{(2l+1)}{4\pi}\frac{(l-|m|)!}{(l+|m|)!}} P_{l|m|}(\cos\theta) \; ,
\end{equation*}
which obeys
\be
\label{prop_sph_func}
\int d\hat{\vec{p}}  \; Y^m_l(\hat{\vec{p\;}})Y^{m'*}_{l'}(\hat{\vec{p\;}})
=  \delta_{ll'} \delta_{mm'} \; .
\ee
One finds the scalar product of these states from \eqref{psi_psi'}:
%\begin{align}
%    \braket{l'm',\mathcal{P}_{\mu}',\beta'|lm,\mathcal{P}_{\mu},\beta} &= \frac%{1}{4\pi} \int d\hat{\vec{p}}\; \big(4u_{1\beta}u_{2\beta}(u_{1\beta}+u_{2\beta%})\big)(2\pi)^6 \delta^{(4)}(\mathcal{P}_{\mu}' - \mathcal{P}_{\mu})\delta_{\be%ta\beta'} Y^m_l(\hat{\vec{p\;}})Y^{m'*}_{l'}(\hat{\vec{p\;}})  ,
%\end{align}
%Finally,
\begin{align*}
  \braket{l', m',\mathcal{P}^{\mu \, \prime},\beta'|l,m,\mathcal{P}^{\mu},\beta}
  =  \; 4\pi u_{1\beta}u_{2\beta}(u_{1\beta}+u_{2\beta})\cdot (2\pi)^4 \delta^{(4)}
  (\mathcal{P}^{\mu \, \prime} - \mathcal{P}^{\mu}) \delta_{ll'} \delta_{mm'} \delta_{\beta\beta'}.
\end{align*}
Thus, the decomposition of the unit operator reads
\begin{equation}
\label{unity}
    \mathbb{1} = \int
    d^{4}\mathcal{P}\sum_{l,m, \beta} \ket{ l,m,\mathcal{P}^{\mu},\beta} \bra{l,m,\mathcal{P}^{\mu},\beta} 
    \frac{1}{N(\beta)} + \ldots,
\end{equation}
where summation runs over all two-particle  states and
\begin{equation}
\label{N}
    N(\beta) \equiv
    2(2\pi)^{5}u_{1\beta}
    u_{2\beta}(u_{1\beta} +
    u_{2\beta}).
\end{equation}
Dots in \eqref{unity}
stand for terms with multiparticle states. 
We  omit these terms in what follows and
comment later on how these terms affect the unitarity relation.

Let us now write the partial wave amplitude,
%which is nothing
%but the transition between states with well-defined  orbital angular momentum 
%$l$
\begin{equation*}
  T^{(l)}_{m' \beta' ; {m} {\beta}} = \braket{l,m',\mathcal{P}^{\mu \, \prime},
    \beta' |T|l, m,\mathcal{P}^{\mu}, \beta} \; .
\end{equation*}
%where the quantum numbers that refer to the initial state are barred.
%This element is
It is given by
\begin{align*}
\label{full_T}
    T^{(l)}_{m'\beta'; {m} {\beta}} &= \frac{1}{4\pi}\int d\hat{\vec{p}}\int d\hat{\vec{p}}\;'\; Y^m_l(\hat{\vec{p}}\;)Y^{ {m^\prime} \, *}_{l}(\hat{\vec{p}}\;')    \braket{\psi\;',\beta'|T|\psi, {\beta}} \; .
\end{align*}
%where $\hat{\vec{p}}$
%and $\hat{\vec{p}}\;'$
%are solid angles, which correspond to the states $\ket{\psi, {\beta}}$ and
%$\ket{\psi\;',\beta}$, respectively
Due to rotational invariance, the $T$-matrix does not vanish only for
$m'=m$ and does not 
depend on $m$~\cite{martin.290916.1,Oller:2018zts}.
Thus, we can write
\begin{align*}
     T^{(l)}_{m' \beta'; {m} {\beta}} &=\delta_{m' {m}}
     \sum_{\tilde{m} = -l}^{ l}
     \frac{T^{(l)}_{\tilde{m}\beta';\tilde{m} {\beta}}}{2l+1}.
\end{align*}
We recall that
\begin{equation*}
  \sum_{{m} = -l}^{ l} Y^{m\, *}_{l}(\hat{\vec{p}}\;') Y^{m}_{l} (\hat{\vec{p}})=
  \frac{2l + 1}{4\pi}
  P_{l} (\cos \gamma) \; ,
\end{equation*}
where  $\gamma \equiv \angle\Big(\hat{\vec{p}}\;',\hat{\vec{p}}\Big)$
is angle between the two momenta, and
 arrive at
\begin{equation*}
T^{(l)}_{m' \beta'; {m} {\beta}}=\frac{\delta_{m' {m}}}{16\pi^2}\int d\hat{\vec{p}}\int d\hat{\vec{p}}\;'\;
P_{l}(\cos \gamma)\braket{\psi',\beta'\;|T|\psi,\beta\;},
\end{equation*}
where, again due to rotational invariance,
$\braket{\psi',\beta'\;|T|\psi,\beta\;}$ does not
depend  on angular variables except for $\gamma$.
Because of this property, it is straightforward to integrate over all
angles but $\gamma$ and obtain
%Changing the variables, one can write
%\begin{equation}
%\label{full_T_no_spin}
%    T^{(l)}_{m\beta; {m} {\beta}} =
%    \frac{\delta_{m {m}}}{16\pi^2}\int d\hat{\vec{p}}\int d\hat{\vec{\gamma}}\;
%P_{l}(cos(\gamma))\braket{\psi',\beta'\;|T|\psi,\beta\;},    
%\end{equation}
%where $\hat{\vec{\gamma}} \equiv\hat{\vec{p}}\;' -\hat{\vec{p}}$.    
%Next, after integration we arrive to 
\begin{equation*}
T^{(l)}_{m'\beta'; {m} {\beta}} =\frac{\delta_{m' {m}}}{2}
\int d(\cos\gamma) \cdot    P_l(\cos\gamma)
\braket{\psi',\beta'\;|T|\psi,\beta\;}.
\end{equation*}
Using
\eqref{matrix_element} one obtains
\begin{equation*}
    \label{T_bb}
    T^{(l)}_{m'\beta'; {m} {\beta}} =  (2\pi)^4\delta^{(4)}(\mathcal{P}^{\mu \, \prime}
    - \mathcal{P}^{\mu})\frac{\delta_{m' {m}}}{2}
    \int d (\cos\gamma) \cdot 
     P_l(\cos\gamma)  M_{\beta' {\beta}}\; .
\end{equation*}
Finally, one defines the partial wave amplitude,
\begin{equation}
\label{a_l}
a_{l,\beta' \beta} =  \frac{1}{32\pi}\int d(\cos \gamma) \cdot
P_l(\cos \gamma ) M_{\beta' \beta},
\end{equation}
and finds
\be
T^{(l)}_{m'\beta'; {m} {\beta}} =
16 \pi \cdot (2\pi)^4\delta^{(4)}(\mathcal{P}^{\mu \, \prime}
  - \mathcal{P}^{\mu}) \delta_{m' {m}} ~a_{l,\beta' \beta}
  \; .
\label{may25-22-10}
  \ee
  
Now we turn to the unitarity relation. Unitarity of $S$-matrix,
%\begin{equation}
$    S S^{\dagger} = S^{\dagger}S = 1$ implies
%\end{equation}
%and for $T$-matrix it is \cite{scattering}
\begin{equation*}
    T-T^{\dagger} = iTT^{\dagger} = iT^{\dagger} T \; .
\end{equation*}
Inserting unit operator given by \eqref{unity} in the right-hand side,
we find
\be
-i \left( T^{(l)}_{m'\beta'; {m} {\beta}} -  T^{(l)\, *}_{m \beta; {m'} {\beta'}}\right)
= \int d^{4}\mathcal{P}''
  \sum_{m'',\beta''} \frac{1}{N(\beta'')}
  T^{(l)}_{m' \beta' ; m''\beta''} T^{(l)\, *}_{m \beta ; m''\beta''}.\nonumber
  \ee
  One makes use of \eqref{may25-22-10} and recalls the definition of
$N(\beta)$, eq.~\eqref{N}, to obtain the unitarity relation in terms of
PWAs:
\begin{align*}
%\label{disting_particles-a}
-\frac{i}{2}\left(   a_{l,\alpha \beta} -  a^*_{l, \beta \alpha} \right)
=  \sum_{\gamma}
  \frac{2}{u_{1\gamma}u_{2\gamma}(u_{1\gamma}+u_{2\gamma})}
  a_{l,\alpha \gamma} a^*_{l,\beta \gamma} ,
\end{align*}
where $u_{1\gamma}$ and $u_{2\gamma}$ are sound speeds of particles in the intermediate state $\gamma$.

One often  assumes time reversal invariance, which gives
$T^{(l)}_{m'\beta'; {m} {\beta}}= T^{(l)}_{m\beta; {m'} {\beta'}}$
and hence  $a_{l,\alpha \beta} =  a_{l, \beta \alpha}$
(see  Appendix A
and Refs.~\cite{martin.290916.1,Oller:2018zts}). In that case the unitarity
relation reads
%\begin{equation}
%  2 \text{Im} T^{(l)}_{m'\beta'; {m} {\beta}} =
%  \braket{l,m',\mathcal{P}^{\mu\, \prime},\beta'|TT^{\dagger}
%    |l, {m},\mathcal{P}^{\mu}, {\beta}} \; .
%\end{equation}
%where we also assume that time-reversal invariance is a symmetry, which implies% that the PWAs are symmetric.
% We prove this property \footnote{This is also demonstrated in \cite{martin.290%916.1,Oller:2018zts}.} in Appendix A. 
%Inserting unit operator given by \eqref{unity} in the right-hand side
%we find
%\begin{equation}
%  2 \text{Im} T^{(l)}_{m' \beta'; {m} {\beta}} = \int d^{4}\mathcal{P}''
%  \sum_{m'',\beta''} \frac{1}{N(\beta'')}
%  T^{(l)}_{m' \beta' ; m''\beta''} T^{(l)\, *}_{m''\beta;'' {m} {\beta}}.
%\end{equation}
%One makes use of \eqref{may25-22-10} and recalls the definition of
%$N(\beta)$, eq.~\eqref{N}, to obtain the unitarity relation in terms of
%PWAs:
\begin{align*}
%\label{disting_particles}
  \text{Im}~   a_{l,\alpha \beta}    =  \sum_{\gamma}
  \frac{2}{u_{1\gamma}u_{2\gamma}(u_{1\gamma}+u_{2\gamma})}
  a_{l,\alpha \gamma} a^*_{l,\gamma \beta} .
\end{align*}
%where we simplified the notations;
%indices 5 and 6 correspond to intermediate state $\gamma$.
For $u_{1\gamma} = u_{2\gamma} = 1$ this relation coincides with
the standard one, see, e.g., Refs.~\cite{Oller:2019opk,DeCurtis:2003zt}.

\subsection{Identical particles}
\label{sec:identical}

We now consider the case of 
identical particles. We again define
 two-particle states as follows:
\begin{equation*}
    \ket{\psi,\beta} = \sqrt{2E_{p_1}}\sqrt{2E_{p_2}} a^{\dagger}_{p_1}a^{\dagger}_{p_2}\ket{0},
\end{equation*}
where  $\beta = \{\phi_i , \phi_i \}$,
%and
%\begin{equation}
%    \bra{\psi',\beta'} = \bra{0}\sqrt{2E_{p_1'}}\sqrt{2E_{p_2'}} a_{p_1'}a_{p_2%'},
%\end{equation}
while the commutational relation is still
given by \eqref{comm_rel}.
%The $u_{\beta}\equiv u_i$
%is sound speed of the each particle in the pair
% $\beta = \{\phi_i , \phi_i \}$.
In the case of identical particles the normalization of the 
two-particle state is different from \eqref{norm_2_part}:
\begin{align}
\label{norm_2_part_id}
&\braket{\psi',\beta'\;|\psi,\beta\;}%= \sqrt{2E_{p_1}}\sqrt{2E_{p_2}}\sqrt{2E_{p_1'}}\sqrt{2E_{p_2'}}\braket{0|a_{p_1'}a_{p_2'}a^{\dagger}_{p_1}a^{\dagger}_{p_2}|0}
\nonumber\\
    &~~~= (2\pi)^6 2E_{p_1}2E_{p_2}\Big(\delta^{(3)}(\vec{p}_1\;'-\vec{p}_1)\delta^{(3)}(\vec{p}_2\;'-\vec{p}_2) + \delta^{(3)}(\vec{p}_2\;'-\vec{p}_1)\delta^{(3)}(\vec{p}_1\;'-\vec{p}_2\;)  \Big)\delta_{\beta\beta'}.
\end{align}
We proceed along the same lines
as in Sec.~\ref{sec:distinguishable}. The change of variables in
\eqref{norm_2_part_id}
gives
\begin{align*}
    %\label{norm_ident}
    \braket{\psi',\beta'\;|\psi,\beta\;} %&= (2\pi)^6 (2u_{\beta}|\vec{p}_1|2u_{\beta}|\vec{p}_2|)\Big(\delta^{(3)}(\vec{p}_1\;'-\vec{p}_1)\delta^{(3)}(\vec{p}_2\;'-\vec{p}_2) \nonumber\\
    %&+ \delta^{(3)}(\vec{p}_2\;'-\vec{p}_1)\delta^{(3)}(\vec{p}_1\;' - \vec{p}_2\;)  \Big) \delta_{\beta\beta'} \nonumber\\
    =(2\pi)^6 \cdot 8u_{\beta}^3 \cdot\delta^{(4)}(\mathcal{P}^{\mu\, \prime} - \mathcal{P}^{\mu})\Big(\delta^{(2)}(\hat{\vec{p}}\;'-\hat{\vec{p}})+\delta^{(2)}(\hat{\vec{p}}\;'+\hat{\vec{p}})\Big)\delta_{\beta\beta'},
\end{align*}
where $u_\beta \equiv u_i$ is the sound speed of the particle $\phi_i$.
The states of definite angular momentum are still given by
\eqref{LSJ_state}, but the scalar product of these states is now %given by
%we want to find $\braket{l'm',\mathcal{P}^{\mu\,\prime},\beta'|lm,\mathca%l{P}^{\mu},\beta}$ in the case of identical 
%particles in the pair. 
%Since we have two terms in brackets in \eqref{norm_ident} we arrive to
\begin{align}
  \braket{l', m',\mathcal{P}^{\mu\,\prime},\beta'|l,m,\mathcal{P}^{\mu},\beta} 
  = \frac{1}{4\pi} \int d\hat{\vec{p}}\;(2\pi)^2 & \cdot 8u_{\beta}^3 \cdot
  (2\pi)^4 \delta^{(4)}(\mathcal{P}^{\mu\,\prime} - \mathcal{P}^{\mu})\delta_{\beta\beta'} \nonumber\\
    &\times
  \Big(Y^m_l(\hat{\vec{p}})Y^{m'*}_{l'}(\hat{\vec{p}}) + Y^m_l(\hat{\vec{p}})Y^{m'*}_{l'}(-\hat{\vec{p}})\Big) \; .
  \label{may27-22-1}
\end{align}
Since identical scalars always have  even\footnote{This
  can be seen also from
  eq.~\eqref{may27-22-1}: the integral in the right hand side vanishes for odd $l$.}  $l$,
we consider even $l$ until the end of this subsection without mentioning this explicitly.  
Making  use of the properties of the spherical functions, eqs.~\eqref{prop_sph_func}, and
\eqref{app:prop_sph_func_1}, we get
%\begin{align}
%    \braket{l'm',\mathcal{P}^{\mu\,\prime},\beta'|lm,\mathcal{P}^{\mu},\beta} &=  \;\pi \big(8u_{\beta}^3\bi%g)(2\pi)^4 \delta^{(4)}(\mathcal{P}^{\mu\,\prime} - \mathcal{P}^{\mu})\Big(1+(-1)^l\Big) \delta_{ll'} \delta%_{mm'} \delta_{\beta\beta'}.
%\end{align}
%For bosons case, $l$ is even number, so
\begin{align*}
  \braket{l', m',\mathcal{P}^{\mu\,\prime},\beta'|l,m,\mathcal{P}^{\mu},\beta} &=  \;2\pi \cdot
  8u_{\beta}^3\cdot (2\pi)^4 \delta^{(4)}(\mathcal{P}^{\mu\,\prime} - \mathcal{P}^{\mu})
  \delta_{ll'} \delta_{mm'} \delta_{\beta\beta'},
\end{align*}
and the contribution of two-particle states with identical particles into the
decomposition of the unit operator reads
\begin{equation*}
     \label{unity_ind}
     \mathbb{1} = \int
    d^{4}\mathcal{P}\sum_{l,m, \beta} \ket{ l,m,\mathcal{P}^{\mu},\beta} \bra{l,m,\mathcal{P}^{\mu},\beta} 
    \frac{1}{N_{identical}(\beta)} + \ldots \; ,
\end{equation*}
where
%summation runs over all two-particle states, and dots mean terms with multiparticle states. For now, as in previous subsection, we omit these terms and comment later on how these terms affect the unitarity relation. The
$N_{identical}(\beta)$ is given by
\begin{equation*}
    N_{identical}(\beta)\equiv
    (2\pi)^{5} \cdot 8u_{\beta}^3 \; .
\end{equation*}
Note that  $N_{identical}(u_{\beta}) = 2 N (u_{1\beta},u_{2\beta})|_{u_{1\beta} =  u_{2\beta} = u_{\beta}}$,
where $N$ has been introduced in \eqref{N}, i.e., if all particles have the same sound speed, then
the normalization factor $N$ is twice larger for identical particles as compared to distinguishable
particles.
%This is consistent with Refs.\cite{}.
%\marginpar{\bf give refs.}
We repeat the calculations in Sec.~\ref{sec:distinguishable} and find that the contribution to PWA
unitarity relation from intermediate states $\gamma$ with two identical particles is given by
%, we can immediately write unitarity relation in the case of identical particles and it reads
\be
-\frac{i}{2}\left(   a_{l,\alpha \beta} -  a^*_{l, \beta \alpha} \right) =
  \sum_{\gamma}
  \frac{1}{2 u_{\gamma}^3}
  a_{l,\alpha \gamma} a^*_{l,\beta \gamma} + \dots \; .
  %\label{identical_particles-a}
  \nonumber
  \ee
  In a $T$-invariant theory one has
\begin{align*}
%\label{identical_particles}
  \text{Im}   ~a_{l,\alpha \beta}    =  \sum_{\gamma}
  \frac{1}{2 u_{\gamma}^3}
  a_{l,\alpha \gamma} a^*_{l,\gamma \beta} + \dots \; .
\end{align*}
This is consistent with  Refs.\cite{Gulmez:2016scm,DeCurtis:2003zt}: if 
all particles have the same sound speed, then the
contribution of identical particles in the intermediate state
has extra factor 1/2 as compared to distinguishable
particles.

\subsection{Unitary bound}
\label{subsec:bound}

We combine the results of Secs.~\ref{sec:distinguishable} and \ref{sec:identical} and write
the PWA unitarity relation as follows:
\be
-\frac{i}{2}\left(   a_{l,\alpha \beta} -  a^*_{l, \beta \alpha} \right)
=  \sum_{\gamma}  g_\gamma     a_{l,\alpha \gamma} a^*_{l, \beta \gamma} ,
\label{general_unit_rel}
\ee
%
%\begin{align}
%\label{general_unit_rel}
%    \text{Im}   a_{l,\alpha\beta}  &=  \sum_{\gamma}  g_\gamma     a_{l,\alpha %\gamma} a^*_{l,\gamma \beta} ,
%\end{align}
where
\begin{subequations}
  \label{may27-22-a}
  \begin{align}
    g_\gamma &= \frac{2}{u_{5\gamma}u_{6\gamma}(u_{5\gamma}+u_{6\gamma})} \;  \;\;\; \mbox{distinguishable}
\label{may27-22-2}
\\
g_\gamma &= \frac{1}{2u_{\gamma}^3} \;  \;\;\; ~~~~~~~~~~~~~~~~~~~ \mbox{identical} \; ,
  \label{may27-22-3}
  \end{align}
\end{subequations}
where eqs.~\eqref{may27-22-2} and \eqref{may27-22-3} refer to distinguishable and identical
particles in the two-particle intermediate state, respectively.
We still do not write explicitly contributions due to multiparticle intermediate states.
We note in passing that eq.~\eqref{general_unit_rel} can be written in the matrix form,
\be
-\frac{i}{2}(   a_{l} - a^\dagger_l) =    a_{l} ~g ~a^\dagger_{l} \; ,\nonumber
\ee
where $g$ is the diagonal matrix with matrix elements $g_\gamma$, and other notations are self-evident.

To obtain the unitary bound, %rite our relation \eqref{general_unit_rel} as follows. 
we introduce rescaled amplitudes $\tilde{a}_{l,\alpha \beta}$ via
\begin{equation}
    a_{l,\alpha \beta} = \frac{\tilde{a}_{l,\alpha \beta}}{\sqrt{g_{\alpha}g_\beta }}\; .
\label{jun3-22-10}
\end{equation}
In terms of the rescaled amplitudes we write the unitarity relation \eqref{general_unit_rel} in a
simpler form:
\begin{align}
\label{general_unit_rel_no_coeff}
-\frac{i}{2}\left(   \tilde{a}_{l,\alpha \beta} -
\tilde{a}^*_{l, \beta \alpha} \right) &=  \sum_{\gamma} \tilde{a}_{l,\alpha \gamma} \tilde{a}^*_{l,\beta \gamma} + \sum_M A_{l , \alpha M} A^*_{l ,M \alpha},
\end{align}
or in matrix form
\begin{align}
\label{general_unit_rel_no_coeff-a}
-\frac{i}{2}\left(   \tilde{a}_l -
\tilde{a}_l^\dagger \right) &=
\tilde{a}_{l} \tilde{a}^\dagger_{l} + A_l A^\dagger_l\; ,
\end{align}
where we  restore
the contribution of multiparticle intermediate
states $M$
in the right hand
side  and  denote schematically the (rescaled)
amplitude $2 \to M$ by $A_{l , \alpha M}$.

Now, let us introduce Hermitean matrices
%\footnote{Notations here are chosen
%  to resemble the harmonic oscillator case, although there is no
%  meaningful analogy.}
\begin{subequations}
  \begin{align*}
    P_l &= -\frac{i}{2}(\tilde{a}_l - \tilde{a}^\dagger_l) \; ,
    \\
    Q_l  &= \frac{1}{2}(\tilde{a}_l + \tilde{a}^\dagger_l) \; ,
  \end{align*}
\end{subequations}
so that
\be
\tilde{a}_l = Q_l + i P_l \; .\nonumber
\ee
Then the unitarity relation reads
\be
P_l = P_l^2 + Q_l^2 + A_l A^\dagger_l - i [P, Q] \; .
\label{may29-22-1}
\ee
We now choose the  orthonormal
basis in two-particle state space in such a way that the
Hermitean matrix $P_l$ is diagonal,
\be
P_{l , \alpha \beta} = p_{l , \alpha} \delta_{\alpha \beta} \; .\nonumber
\ee
In other words, this basis consists of those linear combinations of
states with two particles of definite types which are eigenvectors of
$P_l$.
Then the diagonal $\alpha \alpha$-component of eq.~\eqref{may29-22-1}
is (no summation over $\alpha$)
\be
p_{l , \alpha} = p_{l , \alpha}^2 + (Q_l^2)_{\alpha \alpha} +
(A_l A^\dagger_l)_{\alpha \alpha} \; .\nonumber
\ee
Diagonal  elements of matrices $Q_l^2 \equiv Q_l Q_l^\dagger$
and $A_l A^\dagger_l$ are non-negative\footnote{Because, e.g.,
  $0 \leq \braket {\psi^{(\alpha)}| A_l A_l^\dagger |  \psi^{(\alpha)}}
  = (A_l A^\dagger_l)_{\alpha \alpha}$ for $ \psi^{(\alpha)}_\beta
  = \delta_{\alpha \beta}$.}, so that
\be
p_{l , \alpha}^2 - p_{l , \alpha} \leq 0 \; ,\nonumber
\ee
and therefore
\be
0 \leq p_{l , \alpha} \leq 1 \; .\nonumber
\ee
To cast this relation into somewhat
more familiar form, we come back to the unitarity
relation \eqref{general_unit_rel_no_coeff-a}, sandwich it between
an arbitrary vector
$\ket{\psi}$ of unit norm
and  write, still using the basis of eigenvectors of
$P_l$, 
\be
\braket{\psi| \tilde{a}_l \tilde{a}_l^\dagger |\psi} = \sum_\alpha
p_{l , \alpha} |\psi_\alpha|^2
  - \braket{\psi| A_l A_l^\dagger |\psi} \; .\nonumber
  \ee
  This gives
  \be
  \braket{\psi| \tilde{a}_l \tilde{a}_l^\dagger |\psi} \leq 1,\nonumber
  \ee
  for all $\ket{\psi}$,
  and we arrive at the result that
  \be
  \mbox{all~eigenvalues~of}~~  \tilde{a}_l \tilde{a}_l^\dagger
  ~~\mbox{are~not~greater~than}~1\; .
  \label{may29-22-10}
  \ee
  
  Until now we worked in full generality.
To the best of our knowledge,
    previous analyses not only were restricted to unit sound speed,
    but also studied somewhat    
less general
  situation, see, e.g,
  Refs.~\cite{DeCurtis:2003zt,Lee:1977yc,Lee:1977eg,Chanowitz:1985hj}.
  Namely, (i) 
  the matrix $\tilde{a}_{l , \alpha \beta}$ was assumed to be symmetric 
  due to $T$-invariance,
  $\tilde{a}_{l , \alpha \beta} = \tilde{a}_{l , \beta \alpha}$. Then
  $Q_l$ and $P_l$ are 
  its real and imaginary parts, respectively. (ii) One assumed further that
  $P_l$ and $Q_l$ are simultaneously
  diagonalizable. The latter property holds, in particular,
  when there is just one type of particles,
   and also when the contribution
    of multiparticle states is negligible in
    \eqref{general_unit_rel_no_coeff-a}: in the latter case the imaginary
    part of eq.~\eqref{may29-22-1} gives $[P,Q] = 0$. In this situation
  eq.~\eqref{may29-22-10} tells that any eigenvalue $\tilde{a}_{\alpha \alpha}$
  of matrix $\tilde{a}$ obeys $ |\tilde{a}_{\alpha \alpha}| \leq 1$.
  In fact, in this case one obtains slightly stronger
  bound~\cite{Grojean:2007zz}. In the basis of  eigenvectors
    of $\tilde{a}$
  (i.e., common eigenvectors of $Q$ and $P$),
  one writes the diagonal part of the
  unitarity relation \eqref{general_unit_rel_no_coeff} for each
  $\alpha$  (no summation over $\alpha$):
  %real and imaginary parts of matrix. One chooses the basis in the
%space of two-particle states $\{ \alpha \}$
%in such a way that the matrix $\tilde{a}_{\alpha \beta}$ is
%diagonal (recall that matrix $a_{\alpha \beta}$ and hence matrix  $\tilde{a}_{\%alpha \beta}$ are symmetric,
%see Appendix A). With slight abuse of notations we use the same notation  $\tilde{a}_{\alpha \beta}$ for
%the diagonal matrix. Then the off-diagonal part of the unitarity relation is not particularly useful,
%while the diagonal part reads
\begin{equation*}
\label{unit_rel_multi_part}
\text{Im}   ~\tilde{a}_{l,\alpha \alpha}
=  \tilde{a}_{l, \alpha \alpha} \tilde{a}^*_{l, \alpha \alpha}
  + \sum_M {A}_{l,\alpha M} {A}^*_{l,M \alpha} \; .
\end{equation*}
Again, the contribution of multi-particle intermediate states
is
non-negative, so we arrive at the inequality
\begin{equation*}
%\label{unit_rel_multi_part}
  \text{Im}   ~\tilde{a}_{l,\alpha \alpha}  \geq  |\tilde{a}_{l, \alpha \alpha}|^2 \; .
  \end{equation*}
%$\tilde{a}_{l,\tilde{\alpha} \tilde{\delta}} \tilde{a}^*_{l,\tilde{\delta}\tilde{\alpha}}\geq 0$.
This gives
%Now, following \cite{Grojean:2007zz}, we can find unitarity bound. Let us denote $\Delta \equiv \tilde{a}_{l,\tilde{\alpha} \tilde{\delta}} \tilde{a}^*_{l,\tilde{\delta}\tilde{\alpha}}$. The unitarity relation \eqref{unit_rel_multi_part} can be rewritten as
\begin{equation*}
    \left(\text{Im}\; \tilde{a}_{l, \alpha \alpha} -\frac{1}{2}\right)^2 + \left(\text{Re}\;  \tilde{a}_{l, \alpha\alpha}\right)^2 \leq \frac{1}{4} \;, 
\end{equation*}
and, therefore,
\begin{equation}
   | \text{Re}\;\tilde{a}_{l,\alpha \alpha}| \leq \frac{1}{2},
\label{may29-22-11}
\end{equation}
for any eigenvalue of $\tilde{a}$. 

The latter special situation is particularly relevant when it comes to
perturbative unitarity and the estimate of the strong coupling
scale~\cite{Lee:1977yc,Lee:1977eg,Chanowitz:1985hj,Grojean:2007zz}.
In that case the multiparticle intermediate states
 (almost)
always
give contributions to  \eqref{general_unit_rel_no_coeff-a} which are
indeed suppressed by extra powers of the couplings, while the
matrix $\tilde{a}$ is real   at the tree level.
 Perturbative unitarity then requires that the inequality
\eqref{may29-22-11} holds for the tree level amplitudes.
Note, however, that the bounds
\eqref{may29-22-10} and \eqref{may29-22-11} are qualitatively the same even
in this situation.

\section{Example: theory of two real scalar fields}
\label{sec:unit_rel_holds}

In this Section we show explicitly that the
unitarity relation \eqref{general_unit_rel}
holds at the lowest non-trivial order in a model of two real scalar fields
with the Lagrangian
\begin{equation}
\label{lagr_12}
\mathcal{L} = \frac{1}{2} \left(\dot{\phi}_1^2 -
u_1^2(\vec{\nabla} \phi_1)^2\right) +
 \frac{1}{2} \left(
 \dot{\phi}_2^2 - u_2^2(\vec{\nabla}
 \phi_2)^2\right) +  \frac{\lambda_1}{4!}
 \phi_1^4 + \frac{\lambda_2}{4!}\phi_2^4 + \frac{\lambda_3}{4}
   \phi_1^2\phi_2^2 \; ,
\end{equation}
where $u_1$ and $u_2$ are the two sound speeds.
The scalar potential in eq.~\eqref{lagr_12}
%\begin{equation}
%    V(\phi_1,\phi_2) = \frac{\lambda_1\phi_1^4}{4!} + \frac{\lambda_2\phi_2^4}{%4!} + \frac{\lambda_3\phi_1^2\phi_2^2}{4},
%\end{equation}
is a general fourth-order homogeneous polynomial symmetric under the
transformation
$\phi_{1,2} \to -\phi_{1,2}$. In this theory, the PWA matrix
$a_{\alpha \beta}$ is symmetric due to
$T$-invariance, so the unitarity relation is
\be
\mbox{Im} ~a_{l,\alpha \beta} 
=  \sum_{\gamma}  g_\gamma     a_{l,\alpha \gamma} a^*_{l, \gamma \beta} ,
%\label{general_unit_rel_a}
\nonumber
\ee
or in matrix form
\be
\mbox{Im} ~a_{l} 
=  \sum_{\gamma}       a_{l} g a^\dagger_{l},
\label{general_unit_rel_b}
\ee
where elements of the diagonal matrix $g$ are still given by
eq.~\eqref{may27-22-a}.

\begin{figure}[htb!]
\centering
\includegraphics[width=0.8\textwidth]{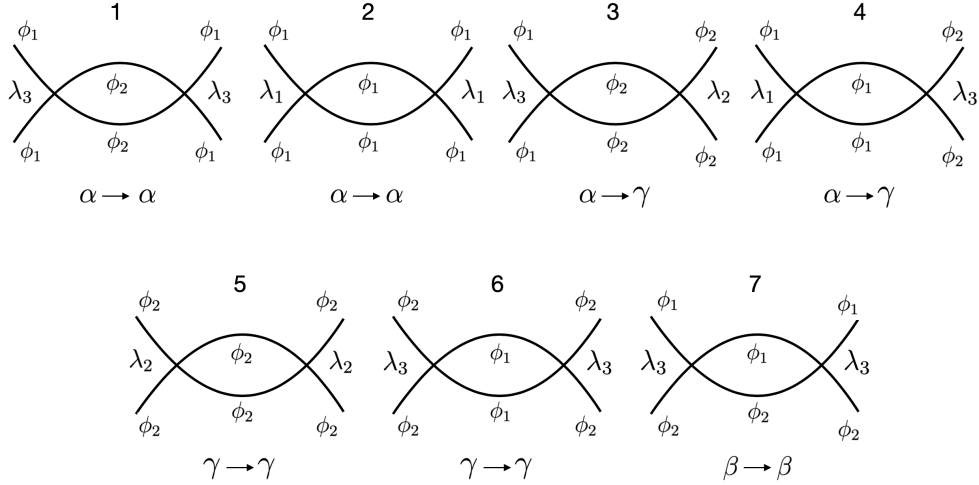}
\caption{ One-loop  $s$-channel diagrams in
  the theory with the Lagrangian \eqref{lagr_12} .}
\label{fig:phi_12_loop}
\end{figure}

The beginning of the calculation follows textbooks. 
There are three
two-particle states $\alpha = (\phi_1,\phi_1)$, $\beta = (\phi_1,\phi_2)$,
and $\gamma = (\phi_2,\phi_2)$ 
in this theory. 
The tree-level matrix elements make a matrix
\begin{align*}
    M_{\text{tree}} = \begin{pmatrix}
M_{\alpha\alpha} & M_{\alpha\beta} & M_{\alpha\gamma}\\
M_{\beta\alpha} & M_{\beta\beta} & M_{\beta\gamma}\\
M_{\gamma\alpha} & M_{\gamma\beta} & M_{\gamma\gamma} 
\end{pmatrix} = \begin{pmatrix}
\lambda_1 & 0 & \lambda_3\\
0 & \lambda_3 & 0\\
\lambda_3 & 0 & \lambda_2 
\end{pmatrix}.
\end{align*}
Since these matrix elements do not
depend on scattering
angle $\gamma$,
the only non-zero PWA, as given by  eq.~\eqref{a_l},  is
$a_0$, 
  i.e., scattering occurs in $s$-wave.
The matrix of these PWAs is given by
\begin{equation}
%\label{a_0_M}
  a_{0 , \text{tree}}
  =   \frac{1}{32\pi}\int^{1}_{-1}d(\text{cos}\theta)P_0(\text{cos}\theta)M_{\text{tree}}
  = \frac{M_{\text{tree}}}{16\pi} =
  \frac{1}{16\pi}\begin{pmatrix}
\lambda_1 & 0 & \lambda_3\\
0 & \lambda_3 & 0\\
\lambda_3 & 0 & \lambda_2
  \end{pmatrix} .
\label{jun2-22-1}
\end{equation}
%Thus, we find
%\begin{align}
%    a_0^{\text{tree}}  = \frac{1}{16\pi}\begin{pmatrix}
%\lambda_1 & 0 & \lambda_3\\
%0 & \lambda_3 & 0\\
%\lambda_3 & 0 & \lambda_2 
%\end{pmatrix}.
%\end{align}
As usual in QFT,
the right-hand side of \eqref{general_unit_rel_b} is of order
$\lambda_i\lambda_j$,  
so Im$\, a_{l}$ obtains its lowest-order contribution
at one loop.  This contribution comes from $s$-channel diagrams
shown in Fig.~\ref{fig:phi_12_loop}, while
 $t$- and $u$-channel diagrams give no contribution to imaginary 
part at  one loop.
%\newpage

We begin
with the first diagram in Fig. \ref{fig:phi_12_loop}.
%Conservation laws in 
%the center of mass frame are
%\begin{subequations}
%\begin{align}
%    &E_1 + E_2 = E_3 + E_4 = E,\\
%    &\vec{p}_1 = -\vec{p}_2,\quad \vec{p}_3 = -\vec{p}_4 \; ,
%\end{align}
%\end{subequations}
%where 
%\begin{align}
%    E_1 =p_1u_1\; , \;\;\;     E_2=p_2u_1 \; , \;\;\; 
%     E_3=p_3 u_1 \; , \;\;\;
%     E_4 = p_4 u_1.
%\end{align}
It gives the one-loop contribution to matrix element 
\begin{align*}
  i  M^{(1)}_{\text{1-loop}}
  %&= \frac{1}{2} (i\lambda_3)^2 \int
  %\frac{d^4q}{(2\pi)^4}\frac{i}{(\frac{E}{2} - q^0)^2-u_2^2\vec{q\;}^2+i
  %  \epsilon}\frac{i}{(\frac{E}{2} + q^0)^2-u_2^2\vec{q\;}^2+i \epsilon}
  %\nonumber\\
    &= \frac{\lambda_3^2}{2}  \int \frac{d^4q}{(2\pi)^4}\frac{1}{\left[(\frac{E}{2} - q^0)^2-u_2^2\vec{q\;}^2+i\epsilon\right]\left[(\frac{E}{2} + q^0)^2-u_2^2\vec{q\;}^2+i \epsilon\right]} \, ,
\end{align*}
where $E$ is still the total energy in
the center-of-mass frame.
Upon rescaling $u_2 \vec{q} \to \vec{q}$, a textbook calculation gives
\begin{equation*}
    \text{Im} \;M^{(1)}_{\text{1-loop}} = \frac{\lambda_3^2}{32\pi u_2^3}.
\end{equation*}
Likewise, the diagrams 2--6
in Fig.~\ref{fig:phi_12_loop} give
%differ from each other only by different 
%constants $\lambda_i$ in vertices and fields on legs and in loop, one can immed%iately 
%write answers for $\text{Im}\;M_{\text{loop},2...6}$:
\begin{subequations}
  \begin{align*}
  & \text{Im} \;M^{(2)}_{\text{1-loop}}= \frac{\lambda_1^2}{32\pi u_1^3}
  \; , \;\;\;
%\end{equation}
%\begin{equation}
  \text{Im} \;M^{(3)}_{\text{1-loop}} = \frac{\lambda_2\lambda_3}{32\pi u_2^3}
  \; ,
  \\
%\end{equation}
%\begin{equation}
   \text{Im} \;M^{(4)}_{\text{1-loop}} &= \frac{\lambda_1\lambda_3}{32\pi u_1^3}
    \; , \;\;\;
%\end{equation}
    %\begin{equation}
 \text{Im} \;M^{(5)}_{\text{1-loop}}
 = \frac{\lambda_2^2}{32\pi u_2^3}   \; , \;\;\;
%\end{equation}
%\begin{equation}
 \text{Im} \;M^{(6)}_{\text{1-loop}}
  = \frac{\lambda_3^2}{32\pi u_1^3}.
\end{align*}
  \end{subequations}
We
now turn to the diagram 7 in Fig. \ref{fig:phi_12_loop}. Unlike others, it
has 
two different particles in the loop.
%The
%conservation laws in this case read:
%\begin{align}
%    &E_1 + E_2 = E = E_3 + E_4,\\
%    &\vec{p}_1 + \vec{p}_2 = \vec{p}_3 + \vec{p}_4 = \vec{k}= 0,\\
%    &\vec{p}_1 = -\vec{p}_2,\quad \vec{p}_3 = -\vec{p}_4,\\
%    &|\vec{p}_1| = |\vec{p}_2|,\quad |\vec{p}_3| = |\vec{p}_4|,
%\end{align}
%\begin{align}
%    E_1 &=p_1u_1,
%    E_2=p_2u_2,\\
%     E_3&=p_3 u_1 ,
%     E_4 = p_4 u_2.
%\end{align}
%One can find that
%\begin{equation}
%    p_1 = p_3 = \frac{E}{u_1+u_2}.
%\end{equation}
%The matrix element is given by
One writes
\begin{align}
\label{m_loop_7_start}
i M^{(7)}_{\text{1-loop}}
%&=  (i\lambda_3)^2 \int \frac{d^4q}{(2\pi)^4}\frac{i}{(\frac{E}{2} - q^0)^2-u_1^2\vec{q\;}^2+i \epsilon_1}\frac{i}{(\frac{E}{2} + q^0)^2-u_2^2\vec{q\;}^2+i \epsilon_2}\nonumber\\
&= \lambda_3^2  \int \frac{d^4q}{(2\pi)^4}
\frac{1}{\left[(\frac{E}{2} - q^0)^2-u_1^2\vec{q\;}^2+i\epsilon\right]
\left[(\frac{E}{2} + q^0)^2-u_2^2\vec{q\;}^2+i \epsilon\right]} \; .
\end{align}
There are four poles of the integrand at
\begin{subequations}
\begin{align*}
    q^0_{1,2} &= \frac{E}{2} \pm u_1 |\vec{q\;}| \mp i \epsilon,\\ 
    q^0_{3,4} &= -\frac{E}{2} \pm u_2 |\vec{q\;}| \mp i\epsilon \; .
\end{align*}
\end{subequations}
\begin{figure}[tb!]
\centering
\includegraphics[width=0.7\textwidth]{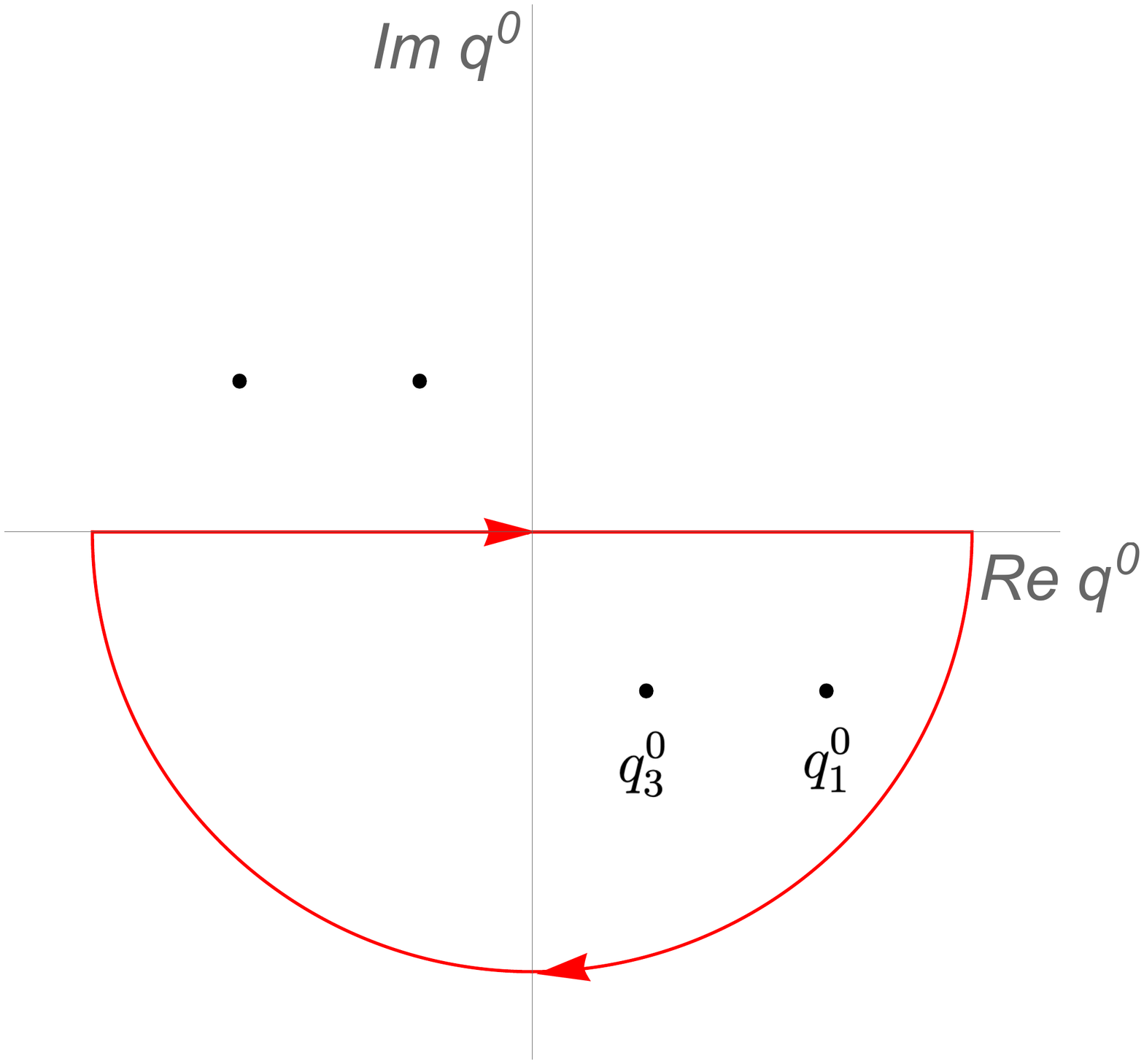}
\caption{Integration contour relevant to
  eq.~\eqref{m_loop_7_start}.}
\label{app:fig:contour}
\end{figure}
Without loss of generality we assume
\be
u_1 \geq u_2 \; .\nonumber
\ee
Then it is convenient to close the integration contour as shown 
in Fig. \ref{app:fig:contour}; the poles inside it are at
  $q^0_{1}$ and   $q^0_{3}$.
 We integrate over $q^0$ and get
%\begin{align}
%    q^0_{1} &= \frac{E}{2} + u_1 |\vec{q\;}| - i\tilde{\epsilon}_1,\\ 
%    q^0_{3} &= -\frac{E}{2} + u_2 |\vec{q\;}| - i\tilde{\epsilon}_2.
%\end{align}
%Taking residuals in these poles we arrive to
\begin{align*}
%\label{im7_res}
i  M^{(7)}_{\text{1-loop}}&= \lambda^2_3\int\frac{d^3q}{(2\pi)^4}(-2\pi i)\Big[\frac{1}{2qu_1
    \big(E+q(u_1-u_2)\big)\big(E+ q(u_1+u_2)\big)} \nonumber\\
    &+ \frac{1}{(-2qu_2)\big(-E+q(u_1+u_2)-i\epsilon \big)\big(E+ q(u_1-u_2)\big)}\Big] ,
\end{align*}
The first term in the integrand does not contribute to   $\mbox{Im}\, M^{(7)}_{\text{1-loop}}$.
Imaginary part due to the second term is calculated using
Sokhotski-Plemelj formula
\begin{equation*}
  \text{lim}_{\epsilon\to 0^+} \left(\frac{1}{x \pm i\epsilon}\right) =
  \mp i \pi \delta(x) + P \left(\frac{1}{x}\right),
\end{equation*}
where $P$ stands for principal value. We find
\be
\mbox{Im}\,  M^{(7)}_{\text{1-loop}} = \lambda^2_3\int\frac{d^3q}{(2\pi)^3}
\frac{1}{2 q u_2\left(E + q(u_1 - u_2) \right)} \cdot \pi \delta \left( -E + q (u_1 + u_2)\right)
  \; ,\nonumber
  \ee
  and, finally,
\begin{align*}
%\label{final_im7}
     \mbox{Im}\,  M^{(7)}_{\text{1-loop}} = \frac{ \lambda^2_3 }{8\pi u_1u_2(u_1+u_2)}.
\end{align*}

To sum up, we collect all results in one matrix
\begin{align}
\label{im_a_loop_2_phis}
\text{Im}\; a_{0 , \text{1-loop}} &= \frac{1}{16 \pi}    \mbox{Im}\,  M_{\text{1-loop}}
%\begin{pmatrix}
%\text{Im}\;a_{\alpha\alpha} & \text{Im}\;a_{\alpha\beta} & \text{Im}\;a_{\alpha\gamma}\\
%\text{Im}\;a_{\beta\alpha} & \text{Im}\;a_{\beta\beta} & \text{Im}\;a_{\beta\gamma}\\
%\text{Im}\;a_{\gamma\alpha} & \text{Im}\;a_{\gamma\beta} & \text{Im}\;a_{\gamma\gamma} 
%\end{pmatrix} 
%\nonumber\\
= \frac{1}{16\pi}\begin{pmatrix}
\frac{\lambda_1^2}{32\pi u_1^3}+\frac{\lambda_3^2}{32\pi u_2^3} & 0 & \frac{\lambda_1\lambda_3}{32\pi u_1^3} +\frac{\lambda_2\lambda_3}{32\pi u_2^3} \\
0 & \frac{ \lambda^2_3 }{8\pi u_1u_2(u_1+u_2)} & 0\\
\frac{\lambda_1\lambda_3}{32\pi u_1^3} +\frac{\lambda_2\lambda_3}{32\pi u_2^3} & 0 & \frac{\lambda_2^2}{32\pi u_2^3}+\frac{\lambda_3^2}{32\pi u_1^3} 
\end{pmatrix} \; .
\end{align}
%Finally, the matrix
%\begin{align}
%    \text{Re}\;a_0  = \frac{1}{16\pi}\begin{pmatrix}
%\lambda_1 & 0 & \lambda_3\\
%0 & \lambda_3 & 0\\
%\lambda_3 & 0 & \lambda_2 
%\end{pmatrix}.
%\label{Re_a}
%\end{align}
Now, eq.~\eqref{may27-22-a} gives for the matrix $g$ in \eqref{general_unit_rel_b}
\be
g = \mbox{diag}\, \left(\frac{1}{2 u_1^3} \, , \frac{2}{u_1 u_2 (u_1+ u_2)} \, , \frac{1}{2 u_2^3}
\right).
\label{jun2-22-2}
\ee
Making use of eqs.~\eqref{jun2-22-1}, \eqref{im_a_loop_2_phis} and \eqref{jun2-22-2}, one
finds that
\be
\text{Im}\; a_{0 , \text{1-loop}} = 
a_{0 , \text{tree}} \, g  \, a_{0 , \text{tree}} \; ,\nonumber
\ee
i.e., the unitarity
relation~\eqref{general_unit_rel_b} is indeed valid to the lowest non-trivial order in the
couplings.

\section{Summary}
In this paper we found PWA unitarity relations \eqref{general_unit_rel} 
in a theory containing  massless
scalar fields with different sound speeds.
We illustrated these relations in a model with the Lagrangian
\eqref{lagr_12}, to the lowest
non-trivial order in the couplings. When written in terms
of rescaled amplitudes \eqref{jun3-22-10},
the unitarity relations have particularly
simple form \eqref{general_unit_rel_no_coeff}, which is formally
the same as in a theory with unit sound speeds.

Using the unitarity relations, we derived the unitarity bounds, which in
the most general case have the form \eqref{may29-22-10}, and
in (still quite general)  case considered in literature reduce to
the familiar form \eqref{may29-22-11} (but written in terms of rescaled
amplitudes). The latter form is particularly useful for evaluating
the quantum strong coupling scale in pertinent EFT.

Our study has been motivated by models with
``strong gravity in the past''~\cite{Ageeva:2021yik}.
One obvious future direction is to make use of our results to
further study models from this class. We anticipate, however, that
the results of this paper may have applications in other
theories where different
perturbations about non-trivial backgrounds
propagate with different sound speeds.

\section*{Acknowledgments}
The authors are grateful to Valery Rubakov for useful comments and fruitful discussions as well as for  careful reading of the early versions of this manuscript. The authors would like to thank Maxim Libanov, Anna Tokareva, Bulat Farkhtdinov, and Sergei Mironov for correspondence and helpful discussions. This work has been supported by Russian Science Foundation Grant No. 19-12-00393.

\section*{Appendix A: Time-reversal invariance and symmetry of $S$ matrix}

In this Appendix we show that $T$-invariance of $S$-matrix implies
the symmetry of the partial-wave amplitudes,
\be
T^{(l)}_{\beta' \beta}  = T^{(l)}_{\beta \beta^\prime} \; .
\label{may24-22-1}
\ee

$T$-invariance of $S$-matrix is invariance under exchange of initial and
  final states and sign reversal of all spatial momenta:
  \be
  \langle \vec{p}^{\,\prime} , \beta'| S | \vec{p} , \beta \rangle =
  \langle - \vec{p} , \beta| S | - \vec{p}^{\, \prime} , \beta' \rangle\nonumber
.  \ee
  We make use of this property to write (we work in the
  center-of-mass frame)
  \begin{align*}
  \langle l, m ; \beta' | S | l,m ; \beta \rangle &=\frac{1}{4\pi} 
  \int~d^3 \hat{\vec{p}}^{\, \prime} ~d^3 \hat{\vec{p}} ~
    Y_{l}^{m \, *} (\hat{\vec{p}}^{\, \prime})  Y_{l}^{m} (\hat{\vec{p}})~
    \langle \vec{p}^{\, \prime} , \beta'| S | \vec{p} , \beta \rangle
    \nonumber \\
    &=\frac{1}{4\pi} 
  \int~d^3 \hat{\vec{p}}^{\, \prime} ~d^3 \hat{\vec{p}}~
    Y_{l}^{m \,  *} (\hat{\vec{p}}^{\, \prime})  Y_{l}^{m} (\hat{\vec{p}})~
    \langle - \vec{p} , \beta| S | - \vec{p}^{\, \prime} , \beta' \rangle
\nonumber \\
    &=\frac{1}{4\pi} 
  \int~d^3 (- \hat{\vec{p}}^{\, \prime}) ~d^3 (-\hat{\vec{p}})~
    Y_{l}^{m \,  *} (-\hat{\vec{p}}^{\, \prime})  Y_{l}^{m} (-\hat{\vec{p}})~
    \langle \vec{p} , \beta| S | \vec{p}^{\, \prime} , \beta' \rangle .
    \end{align*}
    Now, the spherical functions obey
\begin{subequations}
    \begin{align}
    \label{app:prop_sph_func_1}
     Y_{l}^{m} (-\hat{\vec{p}}) &=  (-1)^{l}  \,  Y_{l}^{m} (\hat{\vec{p}}) ,
    \\
    Y_{l}^{m \, *} (\hat{\vec{p}}) &=  (-1)^m  \,  Y_{l}^{-m} (\hat{\vec{p}}),
    \end{align}
    \end{subequations}
      so that
      \be
      Y_{l}^{m \, *} (-\hat{\vec{p}}) =  (-1)^{l+m}\,    Y_{l}^{-m} (\hat{\vec{p}}).\nonumber
      \ee
      This gives
      \begin{align*}
      \langle l m ; \beta' | S | lm ; \beta \rangle
        &=\frac{1}{4\pi} \int
  \int~d^3 \hat{\vec{p}}^{\, \prime} ~d^3 \hat{\vec{p}}~
    Y_{l}^{-m} (\hat{\vec{p}}^{\, \prime})  Y_{l}^{-m \, *} (\hat{\vec{p}})
    \langle  \vec{p} , \beta| S |  \vec{p}^{\, \prime} , \beta' \rangle
\nonumber \\
    & =
    \langle l, -m ; \beta | S | l , -m ; \beta' \rangle .
    \end{align*}
      Since  these matrix elements are
      actually independent of $m$,
    this proves the relation \eqref{may24-22-1}.

\end{document}